# A Polarized View of the Hot and Violent Universe

*A White Paper for the Voyage 2050 long-term plan in the ESA Science Programme*

## Contact Scientist: Paolo Soffitta

*(INAF - Istituto di Astrofisica e Planetologia Spaziali, via Fosso del Cavaliere 100, 00133 Roma, Italy; paolo.soffitta@iaps.inaf.it)*

# INTRODUCTION

Since the birth of X-ray Astronomy, spectacular advances have been seen in the imaging, spectroscopic and timing studies of the hot and violent X-ray Universe, and further leaps forward are expected in the future with the launch e.g. in early 2030s of the ESA L2 mission Athena. A technique, however, is very much lagging behind: polarimetry. In fact, after the measurement, at 2.6 and 5.2 keV, of the 19% average polarization of the Crab Nebula and a tight upper limit of about 1% to the accreting neutron star Scorpius X-1 (not surprisingly, the two brightest X-ray sources in the sky), obtained by OSO-8 in the 70s, no more observations have been performed in the classical X-ray band, even if some interesting results have been obtained in hard X-rays by balloon flights like POGO+ (Chauvin et al., 2017, 2018a,b, 2019) and X-Calibur (Kislat et al., 2018, Krawczynski et al., in prep.), and in soft gamma-rays by INTEGRAL (Forot et al., 2008, Laurent et al., 2011, Gotz et al., 2014), Hitomi (Aharonian et al., 2018) and AstroSAT (Wadawale et al. 2017). The lack of polarimetric measurements implies that we are missing vital physical and geometrical information on many sources which can be provided by the two additional quantities polarimetry can provide: the polarization degree, which is related to the level of asymmetry in the systems (not only in the distribution of the emitting matter, but also in the magnetic or gravitational field), and the polarization angle, which indicates the main orientation of the system.

Fortunately, in 2021 the NASA/ASI mission IXPE (Imaging X-ray Polarimetry Explorer mission, Weisskopf et al., 2016) is going to re-open the X-ray Polarimetry window. IXPE will provide for the first time imaging X-ray polarimetry in the 2 – 8 keV band thanks to its photoelectric polarimeter (Costa et al., 2001), coupled with ~ 25" angular resolution X-ray mirrors (Ramsey et al., 2005). Its orders of magnitude improvement in sensitivity with respect to the OSO-8 Bragg polarimeter implies that scientifically meaningful polarimetric measurements can be done for at least the brightest specimens of most classes of X-ray sources.

IXPE, however, has still an exploratory nature. Polarimetry is a photon hungry technique (the unpolarized part of the signal acts as noise), and with a small class mission like IXPE only the tip of the iceberg can be surveyed. Population studies, especially for relatively faint sources like AGN, are very difficult, and very important science cases like e.g. Gamma-ray Bursts are entirely out of reach.

In 2027, the Chinese-led mission eXTP (Zhang et al., 2019) should also be launched. In addition to timing and spectroscopic instruments, eXTP will have on board photoelectric polarimeters very similar to those of IXPE, but with a total effective area 2-3 times larger. Building on IXPE results, eXTP will iincrease the number of sources for which significant polarimetric measurements could be obtained.

While IXPE will revolutionize the field, and eXTP will consolidate IXPE exploratory results, further progresses are needed to reach a mature phase for X-ray polarimetry. Even if, of course, it is difficult to assess in detail the post-IXPE and eXTP science before their results, some obvious improvements can be easily identified:
1. a broader energy range will allow to study more scientific cases, compare polarization of different spectral components and exploit the diagnostic capability of energy-dependent polarization;
2. a better angular resolution will be crucial in fully exploiting the X-ray polarimetry diagnostic capabilities in e.g. Pulsar Wind Nebulae (PWN) and Supernova Remnants (SNR);
3. a larger effective area will permit, on one hand, to follow the temporal evolution of the polarization properties on a shorter time scale (hopefully down to the relevant physical time scales) and, on the other hand, to perform population studies over statistically significant samples;
4. a wide field polarimeter would allow to catch transients and study the X-ray polarization of the prompt emission, as well as of the afterglows, of Gamma-Ray Bursts.

In the first part of this White Paper we will discuss a few scientific cases in which a next generation X-

ray Polarimetry mission can provide significant advances. While IXPE and eXTP should establish the polarization level of many source classes, the power to extract source physics – from polarization spectral, temporal and spatial variability -- will remain under-explored. In particular, many fundamental questions related to PWN, SNR and even Clusters of Galaxies may be addressed by an imaging X-ray polarimetric mission with a much larger area, better angular resolution and harder energy band than IXPE and eXTP. Such a mission would also provide answers to hot questions related to the physics and morphology of compact objects, both accreting and isolated. Last but not least, the addition of a Wide Field X-ray Polarimetry will allow to perform vital polarimetric measurements of Gamma-Ray Bursts and other transients. In the second part, a possible concept for a medium-class Next Generation X-ray Polarimetry (NGXP) mission will be sketched.

## SCIENTIFIC CASES FOR A NEXT GENERATION X-RAY POLARIMETRY MISSION

### 1. Imaging polarimetry of extended sources

#### *1.1 PWNe*

Pulsar Wind Nebulae (PWNe) are bubbles of relativistic particles and magnetic field that arise as a consequence of the interaction of the relativistic Pulsar Wind confined by its parent Supernova Remnant (SNR) or the ram pressure of the ISM. They are among the most efficient particle (lepton) accelerators in our Galaxy (Nakamura & Shibata 2007, Volpi et al., 2008) and emit synchrotron radiation from Radio to the hard X-ray band (up to MeV energies). They are the only relativistic accelerator where the acceleration site (the pulsar wind shock) is spatially resolved by current optical and X-ray instruments.

The structure of magnetic field in these nebulae has been probed only by Radio, and in a few cases optical, measures. It is found that magnetic field is well-ordered with a polarized fraction close to the theoretical limit. Even if there is no consistent morphological trend, some systems are characterized by a toroidal field geometry (Dodson et al., 2003), and other by a more complex structure (Kothes et al., 2008). The only PWN with a high confidence X-ray polarimetric measurement of P = 19.2 ± 1.0%, is the Crab Nebula (Weisskopf et al., 1978; see also Chauvin et al. 2017).

However, 3D MHD simulations (Porth et al., 2014) predict at the shock front energy equipartition between the particles and the magnetic field, and a large level of instability far away from the injection and acceleration site, suggesting that radio and optical emission are poor tracers of the conditions in the acceleration region, as opposed to X-rays. Moreover, studies with Chandra (Gaensler & Slane 2006) show that PWNe have complex high-energy morphologies with several structures (arcs, rings, jets, tori) with no counterpart at lower energies

Spatially-resolved observations with angular resolution of ≤ 5", would enable us to distinguish and separate these various features from each other, and from the central PSR, in almost all the PWNe of interest (removing also the contamination from the SNR). In particular, the ability to distinguish between the central arcs, the jet and the torus, is pivotal in understanding the physics of acceleration. Different models in-fact predict different relative behaviour among these regions. Shock acceleration based on striped wind reconnection, predict a low level of magnetic energy in the nebula, and potentially a large level of turbulence already at the shock, marked by the internal rings. Fermi-like and shock drift acceleration on the other hand would suggest a large value of polarization in the internal regions. Dissipative acceleration due to turbulence would instead lead to some anti correlation between bright features and polarization, and to strongly depolarized jets. Comparison of polarization properties among various bands from radio to Optical and X-rays will provide the first broad band view of the magnetic field structure in these systems. Below 5" angular resolution, polarization study could benefit from deconvolution, given the typical sizes of the main PWNe.

A collecting area > 5000 cm$^2$ would allow us to either get integrated polarization measures with a Minimum Detectable Polarization at 99% confidence level (MDP99) down to 1% for the 10 brightest nebulae in 1Ms each (3% with 100ks), and an MDP99 down to 10% for a larger set of about 30 PWNe (Kargaltsev & Pavlov, 2008). On the brightest sources like Crab, Vela, MSH15-52, G21.5 it will enable us to get down to a spatially resolved MDP99 10 – 20% in around 100ks. Knowing how the level of turbulence changes with distance from the shock could test recent MHD scenarios invoking the conversion of magnetic energy into particle energy inside the emitting region. Moreover, resolving the jet's magnetic structure with a next generation mission (NGXP) will allow us to test if they are due to PSR precession, that can be related to NS oblateness. This will enable us to investigate polarization in young and old systems, including a few bow-shocks. The ability to cover a wide energy range (0.5 to 20-40 keV) could be used to separate in energy the pulsar contribution both in the case of soft-emitters like Vela or hard one like PSR B1509-58.

The ability to measure and define the polarization direction in PWNe, and its correlation with the observed geometrical axis could also be used to set constraints on Lorentz violation theories that predict rotation of the polarization plane (Machine et al., 2007). Currently existing estimates based on INTEGRAL (Forot et al., 2008) measures in the Crab Nebula are just marginally worse than what derived from optical polarization of GRBs, mostly because of poor determination of the polarization angle in the > 20 keV energy range.

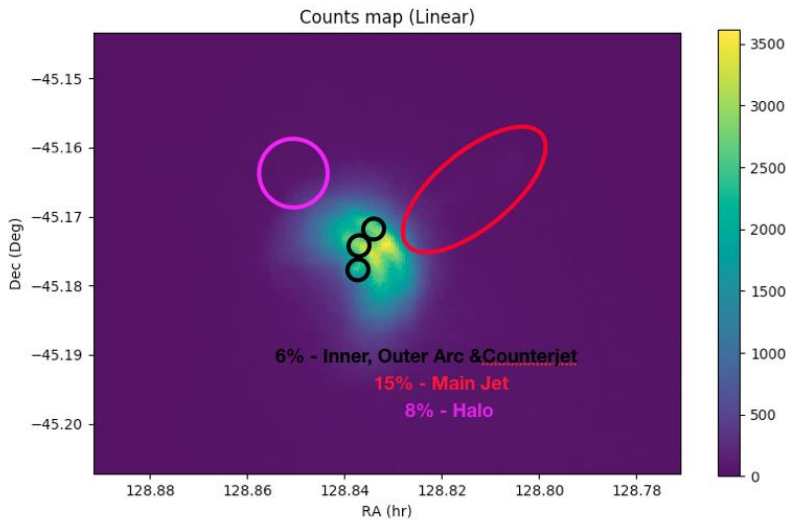

Fig.1.1 Simulated counts map for a 100 ks observation of the Vela PWN, with an instrument with 5000 cm$^2$ effective area and 5" FWHM, in the 2-8 keV energy band. The 2σ minimum detectable polarization of various region (the arcs, jet and halo) is also indicated, assuming a modulation factor 0.3 at 2 keV.

*1.2 SNRs*

X-ray synchrotron polarization studies of supernova remnants (SNRs) are directly related to the question whether SNRs are the main sources of Galactic cosmic rays, and how SNRs are capable to accelerate particles to energies of $3 \times 10^{15}$ eV (the cosmic-ray "knee") or even beyond.

One of the earliest types of evidence that supernova remnants (SNRs) are cosmic-ray accelerators was the identification of their radio emission as synchrotron radiation (Shklovsky 1953) caused by relativistic electrons, with typical energies of $10^9 - 10^{10}$ eV, swirling in 10 – 1000 µG magnetic fields. But this is still 5 orders of magnitude less than the cosmic-ray knee. However, there is a growing body of evidence that SNRs can accelerate particles to much higher energies, although no direct evidence yet that particles are accelerated up to the knee.

The first evidence was the identification of X-ray synchrotron radiation from a subset of young SNRs (Koyama et al., 1995, see Helder et al., 2012 for a review). X-ray synchrotron radiation is caused by $10^{13} - 10^{14}$ eV electrons, much closer to the cosmic-ray knee. These electrons are close to the maximum possible, given that electrons lose their energy much more quickly than protons (years to tens of years) due to the synchrotron radiation itself. It is this X-ray synchrotron radiation that is the target of future X-ray polarization studies. No X-ray polarization of SNRs has ever been detected, but the IXPE mission is expected to lead to the first detection.

The second evidence for very-high energy cosmic rays in SNRs is γ-ray emission, as observed by the

Imaging Atmospheric Cherenkov Telescopes (IACTs). These also provide evidence for the presence of $10^{13} - 10^{14}$ eV particles inside SNRs, but it is not always clear whether these particles are protons and other atomic nuclei (hadronic cosmic rays), or electrons and positrons (so-called leptonic cosmic rays).

Hadronic cosmic rays are considered energetically more important (less than 1% of Galactic cosmic rays are leptons). However, X-ray synchrotron radiation, including polarization studies, provided a lot of diagnostic power for identifying the locations of particle acceleration within the SNRs, and on the conditions of particle acceleration. One reason why X-ray diagnostics was important is of a technical nature: the photon statistics in X-rays exceeds those of gamma-rays, and X-ray imaging can in principle be done on arcsecond angular scales (arcminute scales in gamma-rays). The arcsecond scale corresponds to physical scales of $10^{17}$ cm, similar to the diffusion length scales of $10^{13} - 10^{14}$ eV cosmic rays, but also similar to the length scale over which electrons tend to lose their energy in the post-shock flows (e.g. Helder et al., 2012), so energy dependent polarization maps at fine scale are needed to probe the magnetic field structure in the acceleration zone.

X-ray polarization studies are important as the polarization fraction is a direct measure of how ordered the magnetic fields are. This goes to the heart of the particle acceleration processes (diffusive shock acceleration, DSA, also called Fermi acceleration) itself. In DSA particles gain energy by repeatedly crossing the shock front, due to diffusion in the pre-shock and post-shock regions. If the magnetic field turbulence is high, particles diffuse only slowly, and stay closer to the shock front, allowing for faster acceleration. Faster acceleration also means that particles can be accelerated to higher energies in the limited lifetime during which the SNR has a high shock speed (a few hundred to thousand years). Fast acceleration therefore implies a low polarization fraction. But we also have to assess at what scales the magnetic field is ordered.

Given radio polarization measurements, but also the steepness of the X-ray synchrotron spectra, and the fact that the emitting volumes are smaller for X-ray synchrotron radiation, we expect that X-ray synchrotron emission from SNRs has a polarization fraction between 5-20% (see Vink & Zhou 2018, Bykov et al., 2017, De Ona-Wilhelmi et al., 2017). An important aspect of IXPE will be to establish the overall polarization fraction of a few SNRs, and see if our expectations are correct.

However, the photon statistics with IXPE will be limited, requiring Ms observations. In addition, X-ray polarization will be limited to large regions of the SNRs. For a next generation X-ray polarization mission, one would ideally like to probe the magnetic field topology directly by establishing fluctuations in the polarization fractions, both spatially, on length scales of a few arcseconds, and in the time domain, on time scales corresponding to synchrotron loss time scales of a few years. Since the particle population excite the MHD waves that perturb the magnetic field, dynamical studies of SNR polarization probe the accelerator energetics.

A few arcseconds corresponds to the diffusion length scales of the X-ray synchrotron electrons and also to the magnetic-field fluctuations length scales to which the highest energy protons (~$10^{15}$ eV) are most sensitive (corresponding to their mean free path). Hence, these determine whether the magnetic field fluctuations are large enough to have cosmic-ray acceleration up to the "knee". This will directly probe whether known young SNRs have the right properties to explain the cosmic-ray spectrum observed on Earth.

X-ray synchrotron radiation in Cas A, Tycho's SNR (SN 1572) and Kepler's SNR (SN 1604) are confined to filaments that are arcsecond wide. These filaments can be seen to expand on time scales of years, which in fact corresponds to the shock wave moving through the circumstellar medium. As this happens Alfven waves with different orientations are sweeping through the shock. So, we expect both the polarization fraction and orientation of the magnetic fields to fluctuate over year time scales. The detection of X-ray "twinkling" (Bykov et al., 2009) of small regions should be one of the main aims of a next generation X-ray polarization mission, being a direct result of magnetic field fluctuations.

The requirements on a X-ray mission are quite severe: the effective area should be an order of magnitude larger than IXPE, in order to bring down observations of SNRs to 100 ks, rather than 1 Ms.

Moreover, more and finer pixels are necessary, with a typical pixel resolution of 1" and a half-energy width of < 5".

## *1.3 Clusters of galaxies*

Galaxy clusters are considered as important calibration sources for the nearest generation of X-ray polarimeters since they are bright, extended and not polarized at the sensitivity level accessible with these instruments. However, there is a number of effects that might cause polarization of the X-ray emission from galaxy clusters at lower level. Besides emission from supra-thermal populations of particles in regions with ordered magnetic fields (e.g. connected with cold fronts and shocks, Komarov et al., 2016), these include scattering of the Brightest Cluster Galaxy's quasar emission (intrinsically polarized or non-polarized: Sunyaev, 1982; Gilfanov et al., 1987; Sazonov et al., 2002; Cramphorn et al., 2004) and velocity-sensitive resonant scattering of the line emission (Sazonov et al., 2002; Zhuravleva et al., 2010; Churazov et al., 2010). Probing gas turbulence, magnetic fields evolution and particle kinetics, these signals are capable of providing information on effective collision time scales, viscosity and heat conduction and might help shedding light on fundamental microphysics of the turbulent, weakly-collisional high-$\beta$ plasma of the Intra Cluster Medium (ICM). In the next decades, ICM studies will strongly benefit from data of the next generation X-ray, radio and sub-mm observatories, which will be very well complementary to the sensitive X-ray polarization measurements (e.g. Bykov et al., 2015).

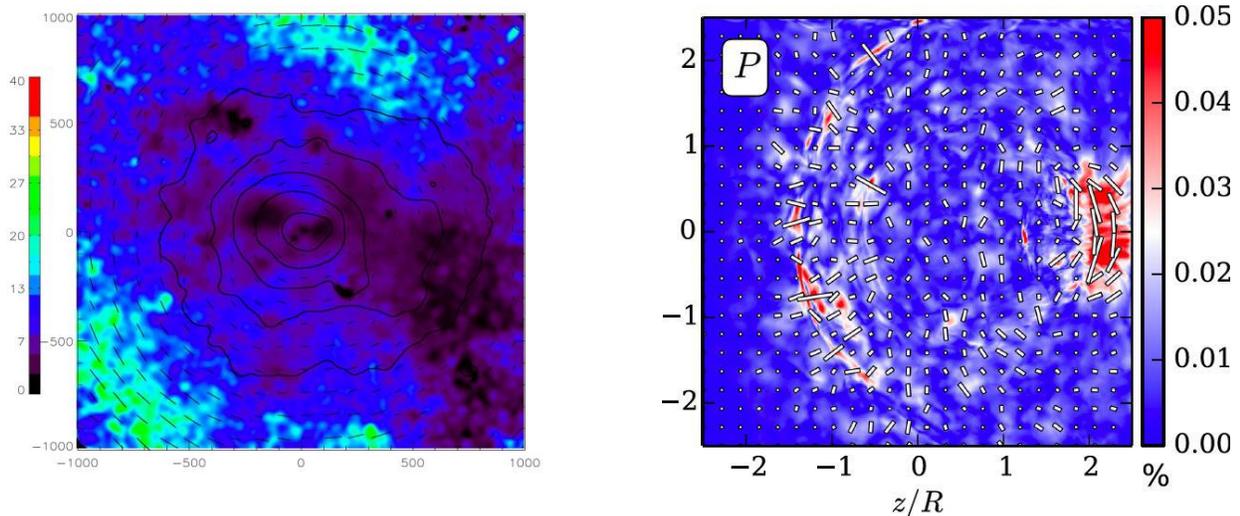

Fig. 1.2: Left: Polarization of the Fe XXV K resonant line emission calculated for a simulated galaxy cluster (2x2 Mpc box) and taking into account turbulent motions inside the ICM (adapted from Zhuravleva et al., 2010). Degree of polarization is color-coded (in percents) and the direction of the polarization plane is shown by the black bars, the black contours show the levels of equal surface brightness spaced by a factor of 4. Right: Polarization of the thermal bremsstrahlung emission from the ICM plasma subject to electron pressure anisotropies in vicinity of a trans-sonic `cold front' substructure (R = 200 kpc) moving inside a 1 Mpc ICM box (adapted from Komarov et al., 2016). Degree of polarization is color-coded (in percents) and the direction of the polarization plane is shown by the white bars.

While measuring intrinsic polarization in galaxy clusters is an interesting and far reaching exercise, the degree of polarization is very small, which makes such observations extremely challenging. At the same time, for a wide range of other problems the clusters of galaxies can be used as a convenient calibration (unpolarized) source and/or for axion searches. Axions and axion-like particles are a generic feature of many extensions to the Standard Model. In fact the QCD axion is the best explanation for the absence of CP violation in the strong force. Radiation polarized parallel to a magnetic field couples to axions, so that initially unpolarized radiation can become polarized as it travels through the magnetic field. Distant galaxy clusters whose bremsstrahlung emission should be nearly unpolarized can provide an excellent probe of axion physics (and the structure of intergalactic magnetic fields) with a future X-ray polarimeter. An observation of polarized X-rays from such a source would indicate that axions exist and provide a constraint on the coherence length and strength of intergalactic magnetic fields combined with the axion coupling. Studying clusters at a range of redshifts would constrain the coherence length, and other observations can constrain the intergalactic magnetic field, yielding constraints on the axion itself. The key to such a test is sensitivity because a smaller upper limit on the observed polarization yields tighter constraints on the axion.

## 2. *X-ray Polarimety of Compact Objects*

### 2.1 Accreting Neutron Stars

Accreting neutron stars come in a number of incarnations: 1) classical strong-field X-ray pulsars (XRPs) mostly in high-mass X-ray binaries; 2) accreting millisecond pulsars in low-mass X-ray binaries and 3) weakly- (or non-) magnetized neutron stars in low-mass X-ray binaries showing no pulsations.

*2.1.1 Classical strong-field pulsars*
Accreting XRPs are among the brightest X-ray sources and represent prime targets for X-ray polarimetry. Typical XRPs are characterized by the magnetic field in the range $B=10^{12}$-$10^{13}$ G as measured by the presence in the X-ray spectra of many objects of broad absorption lines appearing at 11.6 ($B/10^{12}$ G) keV, which are interpreted as cyclotron resonance scattering features (CRSF). In such strong magnetic fields, photons are linearly polarized in two normal modes, the ordinary (O) and extraordinary (X) ones. Below the electron cyclotron energy, the X-mode opacity is highly suppressed with respect to the O-mode one resulting in nearly 100% polarized emerging radiation in the X-mode. Variation of polarization position angle (PA) with pulsar phase immediately determines the geometry of the emission region, the inclination of the pulsars, the magnetic obliquity, and even the PA of the pulsar rotation axis on the sky and the sense of rotation.

Despite several decades of studying, the configuration of the emission region in XRPs remains uncertain. It is generally believed that at low luminosities the accreting matter impinges the neutron star surface close to magnetic poles producing a hotspot that is responsible for emission. Above some critical luminosity the matter is stopped in a radiation-mediated shock above the surface and emission is produced in the accretion column. Such different emission regions should have very different emission patterns (`pencil´ and `fan´ beams). Reconstruction of the intrinsic beam patterns of XRPs based on the observed pulse profile alone is currently strongly model dependent, but polarization can break degeneracies. Observations of XRPs during outbursts that are expected to pass through the critical luminosity would allow to see directly strong changes in polarization properties. High sensitivity of NGXP will allow us to study wind-fed pulsars accreting at sub-critical rates will allow a direct comparison of the accretion region geometry in disc- and wind-fed systems. Furthermore, a high angular resolution will allow to study emission region geometry of ultra-luminous X-ray pulsars in external galaxies.

Apart from this, the sweep in the PA can also act as a tracer of the *B*-field geometry in XRPs, which remains a subject of long-standing debate. Any rapid change of the PA may be an indication of a non-dipolar magnetic field. In particular, the polarimetric observations across the line will allow us to probe deeper into the geometry of the accretion column and magnetic field, to study in great details variation of polarization with super-orbital phase and to test directly free-precession model for Her X-1 and maybe to see this effect in other sources.

*2.1.2 Accreting millisecond pulsars and neutron star equation of state*
Accreting millisecond pulsars (AMPs) contain weakly magnetized neutron stars (with $B=10^8$-$10^9$ G) accreting matter from a typically rather small companion star (e.g. Patruno & Watts 2012; Campana & Di Salvo 2018). About 20 sources, with periods ranging from 1.67 to about 10 ms, are known at present, all of them are transients which go into outbursts every few years. Neutron stars in these systems have been spun up by accretion to the millisecond periods. Similarly to classical XRPs, the accreting matter follows the magnetic field lines hitting the surface close to the magnetic poles, but the emission region is much larger, owing to the fact that magnetosphere is only a few stellar radii. Radiation produced by Compton scattering in the accretion shock is likely polarized at the level 10-20% (Viironen & Poutanen 2004).

The information about the emission pattern, the rotational velocity and the neutron star compactness is recorded into the observed pulse profiles, which can be used to recover neutron star mass and radius and to constrain the neutron star equation of state (e.g. Poutanen & Gierlinski 2003). A degeneracy between the number of parameters can be broken with the polarization data, which will allow to measure independently the pulsar inclination and magnetic obliquity (see Fig.2.1), significantly improving the mass-radius constraints. It is likely that by the time of NGXP, we will know the equation of state rather accurately, and polarimetric data will then be used to study variations in the accretion geometry on short time scales. The energy dependence of the polarization degree (PD) will also allow us to independently measure the temperatures of the electrons in the shock and of seed photons as well as to test Comptonization model for the hard spectral component. A possible energy dependence of the PA can give a clue on the emission region geometry, in particular on a possible displacement of the shock from the blackbody emitting region. A high sensitivity NGXP will also allow to study properties of the reflected component from the accretion disc, its pulse phase variations, and to determine independently disc inclination and the magnetospheric radius.

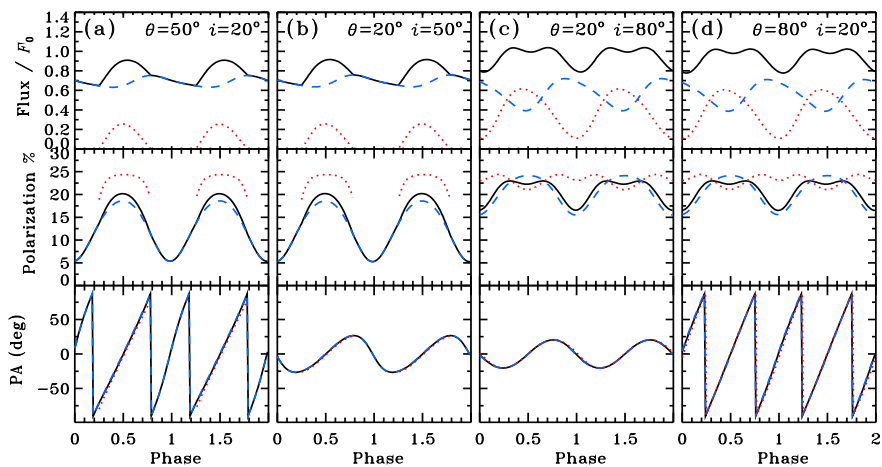

Fig. 2.1: The pulse profile as well as the phase-dependence of the PD and PA. Black solid line give the contribution of two antipodal spots, while the dashed and dotted line correspond to the contribution of each spot separately. The pulse profile and PD are degenerate to exchanging inclination I with the magnetic obliquity θ, while the PA shows dramatically different behaviour allowing both angles to be obtained (adapted from Viironen & Poutanen 2004).

*2.1.3 Weakly magnetized neutron stars*

Most of the neutron stars in low-mass X-ray binaries have too low magnetic field and show no pulsations. The accretion disc extends there all the way to the stellar surface forming a boundary or a spreading layer. At low accretion rates, the inner disc may be replaced by a hot flow, similarly to the hard-state black holes. NGXP operating in the soft X-rays, below 1 keV, would be able to study polarization of the accretion disc radiation and to see the rotation of the PA with energy, the same general-relativity effects as one expects to see in accreting black holes in their soft states. Such measurements may provide information about the otherwise unknown neutron star spin. Radiation from the whole boundary/spreading is only weakly polarized at a 1% level, but its reflection from the accretion disc is much stronger polarized and will be clearly seen in the energy dependence of the PD and will give an estimate of the disc inclination (Lapidus & Sunyaev 1985). Strongly polarized reflected emission is also expected during X-ray bursts from the atoll sources. In the hard-state, atoll sources the X-ray emission is dominated by a hard component, which is likely produced in the inner hot flow. It may be polarized at a few per cent level due to Compton scattering in a flattened geometry (Poutanen & Svensson 1996). One expects strong energy-dependence of the PD, because PD grows with scattering order. Interestingly, if we will also detect rotation of the PA with energy, we will be able to detemine the location of the hard X-ray emitting region much more precisely than is currently known.

## *2.2 Magnetars*

The study of highly magnetized sources is by far one of the key drivers for an X-ray polarimetric mission. Isolated Neutron stars (INSs) are the strongest magnets known in the Universe (with magnetic fields up to the petagauss range) and among them magnetars are their most extreme manifestation. Once restricted to a dozen sources, magnetars are by now known to be present in a variety of astrophysical settings, from gamma ray bursts to Soft Gamma Repeaters and Anomalous X-ray Pulsars. The discovery of "low-field magnetars" has made clear that the strength of the dipolar magnetic field alone (as measured by X-ray timing) is not a determinant of magnetar activity, but it is the topology of the field itself that causes their spectacular phenomenology (see e.g. Turolla et al., 2015, Kaspi & Beloborodov 2017 for recent reviews on magnetars).

The nascent field of broadband X-ray polarimetry offers revolutionary new ways to investigate the physics and geometry of magnetars and other, possibly related, classes of INSs: the X-ray dim isolated neutron stars (XDINSs) and the Central Compact Objects (CCOs), in particular (Taverna et al., 2014, 2015, Gonzalez Caniulef et al., 2015). These sources are radio-silent so that polarization measurements are possible only in X-rays (in a handful of cases maybe also in the optical, Mignani et al., 2017). Mapping super-strong magnetic fields and their formation, evolution and instabilities also allows us to unveil the link among magnetars and the most spectacular fireworks, such as gamma ray bursts, and gravitational waves. Explaining the different NS manifestations, the physics behind them, and the relations among different NS types is one of the most challenging goals in compact objects astrophysics and in this respect broadband X-ray polarimetry with large throughput offers the key to the ultimate understanding of the endpoints of massive star evolution. Constraints on the NS spin and magnetic fields are vital inputs for leading models of GRB central engines and kilonova emission. Furthermore, such a potential mission enables the use of neutron stars as cosmic laboratories to quantitatively test QED high-field predictions (Taverna et al., 2014; see also Heyl & Shaviv 2000, 2002), and to look for axion-like particles (ALPs) (Fortin & Kuver 2019).

After many decades, a space-borne X-ray polarimeter will fly at last with the NASA SMEX mission IXPE (Weisskopf et al., 2016) in 2021, while eXTP (Santangelo et al., 2019) is expected to follow in a few years. These missions will start tackling some of the key physical issues concerning highly-magnetized INSs, in particular by measuring the polarization degree in some bright magnetar sources and performing the first experimental test of QED vacuum birefringence. However, despite their utmost importance as pathfinders, the science impact of both missions is hampered by their relatively small collecting area and, most crucially, restricted energy range. Both experiments are in fact capable to measure X-ray polarization only in the ~ 2 – 10 keV range. The soft X-ray band (below 1 – 2 keV), where the surface of the neutron star itself is visible (and polarimetry can probe the mass-radius relation for neutron stars and QCD), and the high-energy range, where the most interesting magnetospheric processes and the outburst/flaring activity are detectable, will remain unexplored.

Polarimetry is photon-hungry and observations of INSs with currently planned missions are limited by the long exposure times. IXPE, with an effective area of 700 $cm^2$ at 2 keV, typically needs 0.5 – 1 Ms to measure the polarization degree in a magnetar source with the accuracy required to test QED vacuum birefringence. eXTP would improve over this figure but the collecting area of its Polarimetry Focusing Array is 900 $cm^2$, still smaller than that designed for XIPE, the ambitious polarimetric mission submitted (but not selected) for ESA M4. This will limit the number of magnetars to be targeted in the first two years of operations to 2 – 3 at best. In order to have a full-fledged impact on NS science, polarimetry needs an X-ray instrument with a large collecting area (> 2000 $cm^2$) that will cut exposure times down to 0.1 Ms. This will allow observations of 10 – 15 sources. Looking at a variety of sources will reduce the effects of systematics of a peculiar source on probing the underlying physics of QED and QCD while simultaneously uncovering new phenomena that will only be apparent in the polarized signal. Furthermore, an instrument with a large collecting area, together with a polarimeter with extended energy coverage, possibly in the 0.1 – 30 keV range (or even more), will allow to answer some key questions concerning the links among the different INS populations, the physics behind magnetar bursts, the state of NS surfaces and the structure of their magnetic field.

*2.2.1 High-field radio pulsars (HBPSRs) vs. magnetars*
A super-strong magnetic field is not a feature unique to magnetars. "Normal" radio pulsars can host surface dipolar fields well in the magnetar range in the so-called HBPSRs (e.g. Kaspi 2010). However, no sign of magnetar activity has been seen in these sources. The current explanation is that the magnetospheres of HBPSRs do not support currents flowing along the closed field lines, because no magnetic helicity is transferred from the internal to the external field, contrary to magnetars. HBPSRs are detected in the X-rays but they are too faint for IXPE or eXTP. A measurement of their polarization degree, which is expected to be about 30% in magnetars, will unambiguously show if this is indeed the case and will help shedding light of the possible evolutionary links between the two classes.

*2.2.2 Fireballs and bursts*
The emission of short (≈ 0.1 – 10 s), energetic (≈ $10^{36}$ – $10^{43}$ erg/s) bursts of hard X-rays is one of the distinctive characteristics of magnetars (Turolla et al., 2015, Kaspi & Beloborodov 2017). We still do not know where the instability that triggers such spectacular activity is located (in the crust, liquid core or magnetosphere of the NS) but rapid magnetic field reconfiguration is without doubt an integral part of burst emission. Independently on the main engine, gamma rays are thought to propagate in a hot pair fireball which expands from the star surface up to few star radii (Thompson & Duncan 1995). Modelling of the burst spectra indicate that under such conditions the emitted radiation may be either highly polarized in the X mode or present a shift at a certain energy between X and O mode polarization (Taverna & Turolla 2017, Israel et al., 2008). IXPE and eXTP might be able to catch the polarization signal only in some, relatively rare, very bright bursts (the so-called intermediate flares) are at any rate outside the primary science targets. An improved detector, sensitive above 10 keV, will systematically measure the polarization degree also in less energetic, much more common events, putting the fireball scenario to a definite test, identifying different subfamilies of bursts and ultimately nailing down the emission process.

*2.2.3 Phase-resolved spectro-polarimetry*
Because of the scanty statistics, only phase-resolved, energy-integrated polarimetry (or the opposite) of magnetars will be possible with IXPE/eXTP. X-ray observations of some magnetar sources (most notably SGR 0418+5729) have shown the presence of phase- and energy-variable absorption features at a few keVs (Tiengo et al., 2013, Rodriguez Castillo et al., 2016). The nature of these lines is still not entirely clear but, if interpreted in terms of proton resonant absorption/scattering, they provide a direct measure of the strength and a probe of the small-scale structure of the magnetic field close to the surface. Phase- and energy-resolved polarimetry will be within range of a future instrument with large collecting area and will make it possible to directly test the nature of these absorption features.

*2.2.4 Getting the feel of a NS surface*
In the case of thermally-emitting INSs (the XDINSs) the extended energy range of a future mission can scrutinize the polarization in the soft (0.1 – 2 keV) X-ray band. Phase-resolved spectro-polarimetry will provide a measure of the surface magnetic field and of the source geometry. Since the polarization observables are very sensitive to the surface emission model (a condensed surface, an atmosphere), polarization measurements will give for the first time a direct feel of the physical state of the star surface layers, which has not yet been obtained through spectroscopy alone (Gonzalez Caniulef et al., 2016). With sufficient <keV sensitivity we can also study the surface polarization of the brightest classical pulsars. This provides the opportunity to probe the physics of the magnetized neutron star atmosphere and, with phase-resolved studies, to probe the magnetic structures associated with the radio-emitting poles (e.g. Pavlov & Zavlin 2000).

*2.2.5 The spin-velocity alignment and GW emission*
Unless a surrounding PWN shows toroidal structure, X-ray polarization is the main way to probe the spin orientation of isolated, radio quiet pulsars (see fig 2.2). Although radio pulsars can offer clues, because the pulsar emission mechanism itself is poorly understood and perhaps only a fraction of neutron stars are born to be radio pulsars, the radio pulsar population offers only a biased look at this problem, whereas the thermal emission from the neutron-star surface is a generic feature of the cooling ember of the supernova core. The only measurement of optical polarization of a XDINS, RXJ1856 [6], allowed to nail down the direction of the spin and magnetic axis on the plane of the sky, and their inclination with respect to the NS proper motion. The issue of the pulsar-spin velocity alignment, the possibility to have orthogonal spin-velocity configurations is still highly debated and intimately linked to the way in which a velocity kick is transferred to the proto neutron star during the formation and to the duration and physics of the star acceleration phase (Noutsos et al., 2012, Spruit & Phinney 1998). In fact, how neutron stars get their kicks is key to understanding whether supernova can generate detectable bursts of gravitational radiation. If the kicks are sudden, we expect more gravitational radiation and no particular alignment between the spin and proper motion. On the other hand, if the neutron star gradually accelerates, we expect little gravitational emission and alignment between spin and proper motion (Ng & Romani 2007).

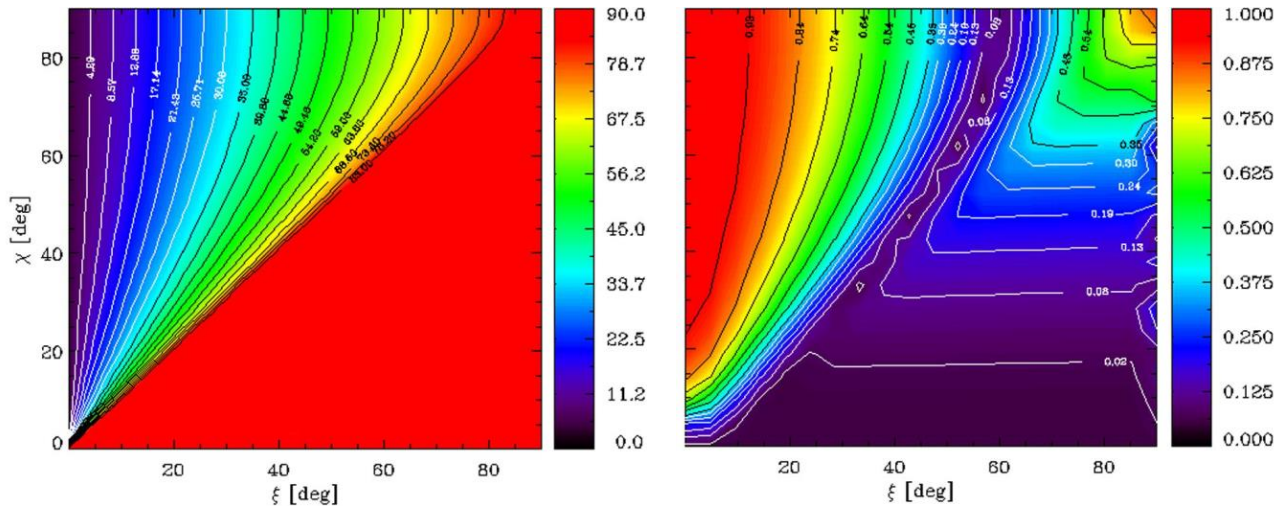

Fig. 2.2: The polarization angle and polarization fraction of neutron star thermal surface radiation in the soft X-rays (0.3 keV) as a function of the two geometrical angles ξ and χ which measure the inclination of the line-of-sight and of the magnetic axis with respect to the star spin axis (from Taverna et al., 2015).

### *2.3 Accreting stellar-mass black holes*

Black holes in X-ray binaries (XRBs) are fed by their orbiting donor star via an accretion disk. In some cases, the accretion in these systems may be persistent, however, in most cases transient accretion events are observed. Currently we know around 77 transient and 10 persistently accreting black holes or black hole candidates (Tetarenko et al., 2016). The majority of the transient sources (63) have undergone only one outburst. These outbursts may last weeks to months, and while they last, the accreting black hole system undergoes huge changes that manifest themselves through radically different spectral properties. These sources repeatedly go through the cycles between two main X-ray spectral states (Remillard & McClintock 2006). In the very X-ray luminous ($10^{38-39}$ erg/s) (Chaty et al., 1996, Taam et al., 1997, Zhang et al., 1997) soft state, a thermal accretion-disk component and a steep power-law tail dominate the spectrum. In the hard state, the spectrum is predominantly a flat power law. During the hard state relativistic radio jets are often present (Fender 2001). X-ray emission origin in these sources may be from an accretion disk, a hot corona, or a relativistic jet. The power-law component is due to Comptonization of the thermal disk emission in a hot corona (Sunyaev & Titarchuk 1980, Hua & Titarchuk 1995) which, in the hard state, may be a compact, magnetized jet (e.g. Kylafis & Belloni 2015). This power law component should be polarized (Poutanen 1994, Sunyaev & Titarchuk 1985, Poutanen & Vilhu 1993), with a polarization degree which depends on the geometry (and thence on the nature and origin) of the corona (e.g. Tamborra et al. 2018). A next generation X-ray polarimetric mission will build on the heritage from the polarimetric pathfinder missions IXPE in the early 2020's by NASA, and the Chinese polarimetric mission eXTP in the late 2020's and early 2030's. These missions should allow us to validate and improve our methods to measure black holes spins, and should give us new insights about the shape and location of the accretion disk coronas and the origin of the ~ 3–8 keV excess emission currently believed to be relativistically broadened Fe K-alpha fluorescence mostly coming from the immediate surrounding of the black hole horizon. A polarimetric mission with a much larger effective area, better resolution in energy and broader energy band coverage will provide us with much more detailed information and will enable us for the first time to do true timing spectro-polarimetry. Thus, we can expect new advances in the following areas, as such a mission will be able to give high signal-to-noise time resolved results for all stellar mass black holes exhibiting fluxes exceeding a few mCrab.

*2.3.1 Evolution of spectral transitions via X-ray polarimetry*
With the NGXP mission we would be able to track the evolution of the polarization properties throughout the whole cycle of XRB spectral transitions. Here, the high energy polarization detector will be important for study of the XRB in their hard state, providing us with valuable information on emission from the base of the jet and/or the hot corona. The high energy detector together with the increased energy resolution in the soft energy band will also enable us to measure the spin of the black hole in the soft state of XRBs with greater robustness and higher precision.

*2.3.2 Black hole spin measurements in the soft state of XRBs*
Stellar-mass black holes are the most promising sources to probe strong gravity effects and to put constraints on the angular momentum (spin) of black holes. They provide us with large X-ray count rates and the cleanest method to measure the spin when in the so-called high/soft state. In this state, the dominant spectral component in the soft energy band is thermal emission from the accretion disk. Due to strong gravity effects, the observed polarization properties change with energy (Connors et al., 1980, Matt 1993, Dovčiak et al., 2008, 2011, Schnittman & Krolik 2009, 2010). Due to self-irradiation of the disk, the reflected thermal component contributes mainly at higher energy with the polarization angle perpendicular to the one originating from the direct thermal component. Thus one expects a change in the polarization angle at a certain energy. Since this transition energy depends on the BH spin, its precise measurement with a high-resolution detector will give us more robust and more precise BH spin values, see Fig.2.3. The transition for low spins happens in the tail of the thermal emission (thus at low flux and high energy) thus only the high sensitivity of NGXP together with its hard X-ray polarization detector will enable us to observe it and thus improve our BH spin measurements. The hard X-ray polarization detector will also measure possible "contamination" by the hard spectral component (see below) from the corona (or failed jet). Such broadband measurements will be able to disentangle the polarization of the two components.

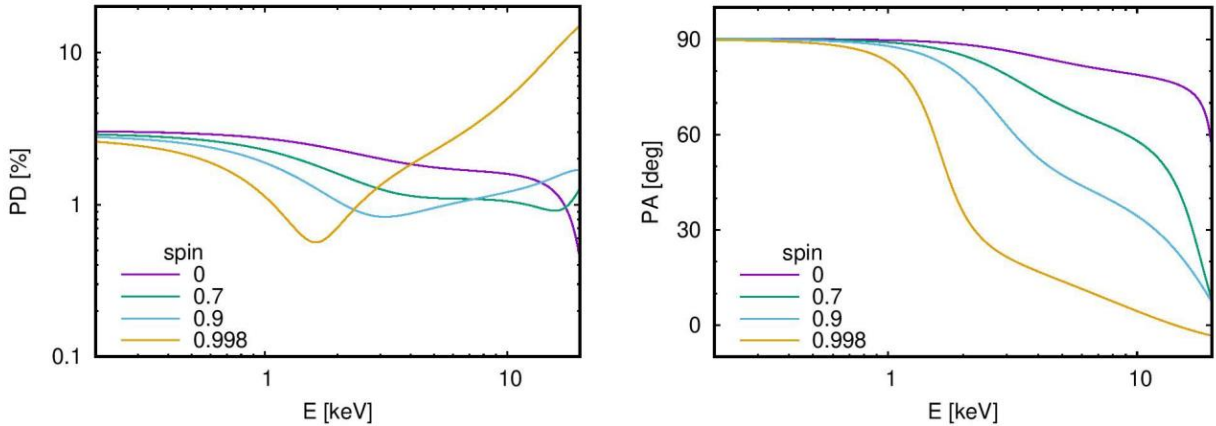

Fig. 2.3: Polarization degree (left panel) and polarization angle (right panel) dependence on energy. Note the dip in PD at the energy where PA changes due to the contribution of reflected self-irradiation of the disk. From Taverna et al., in prep.

*2.3.3 Generation of jets and connection of jet with BH spin*
By the time of NGXP, we hope that the long-standing puzzle about the nature of the corona in XRB will be solved. If IXPE and/or eXTP confirm that the main contributor to the hard X-ray power-law emission In the hard state is a compact magnetized jet (Kylafis & Belloni 2015), its polarization may be significantly larger than in case of different, less asymmetric configurations. With this mission, we will additionally be able to observe the evolution of the polarization degree and angle with time. This will give us further clues on how these jets are formed. We will measure the evolution of the magnitude and structure of the magnetic fields near the black holes and study the jet connection (e.g. its power) with the spin of the black holes. The high mass black hole XRB system Cygnus X-1 may provide one of the best targets for studying the origin of jets. The micro-quasar SS443 is another very interesting source that features spectacular, large-scale, bi-conical jets.

*2.3.4 Evolution of the corona.*
If, on the other hand, it is confirmed that the primary X-ray spectrum of X-ray binaries is produced in a hot corona surrounding the compact object by Comptonizing soft photons from the accretion disk on very energetic electrons, we will be able to study the properties of corona - its shape, size and position above the accretion disk, its optical thickness and electron temperature. With the large effective area of the NGXP we will also be able to observe how the corona changes with time as the hard state evolves from quiescence to peak luminosity and further to the soft state.

*2.3.5 Accretion disk dynamics: Quasi-Periodic Oscillations (QPOs) and Bardeen-Petterson (BP) effect*
NGXP will push our understanding of black hole accretion to the next level by enabling time-resolved studies of the *accretion disk dynamics*. The combination of the General Relativistic Lense-Thirring (LT) precession of matter orbiting a Kerr black hole close to its event horizon and the disk turbulence is expected to lead either to the precession of the central disk as a solid body, to warped disks, or to even more interesting unstable configurations. The LT precession of the inner disk is one of the favoured models to explain the Low Frequency Quasi-Periodic Oscillations (LFQPOs) in the range of ~0.1-10 Hz that are regularly observed in X-ray binaries (Ingram et al., 2009). The model predicts the right range of observed LFQPO frequencies. The model predicts a modulation of the polarization fraction and angle (Ingram et al., 2015) which could easily be detected with the NGXP and used to reconstruct the 3-dimensional orientation of the black hole spin axis at the centre of the precession cone. The wider bandpass of the mission would furthermore enable the detection and characterization of warped disks and disk instabilities (Cheng et al., 2016, Abarr & Krawczynski, in prep.)

*2.3.6 Contribution of the companion to the polarization signal*
Part of the X-ray emission originating in the accretion disk, corona or jet is reflected from the companion in close high mass XRBs inducing a polarized component in the source spectrum that would be orbital phase dependent. This will also be a good diagnostic for the inclination of the orbit with respect to the line of sight, with a number of implications for the knowledge of the binary system.

## *2.4 Gamma-ray Bursts*

Recently, gravitational waves (GW) from merging black holes have been detected, short Gamma Ray Bursts (GRB) have been proven to be produced by GW-emitting compact binary stellar mergers with accompanying optical kilonovae and very high energy TeV emission from a long GRB has been detected for the first time. These herald a new era of multi-messenger black-hole astronomy. Rapid-response observations of black-hole driven transients probe regions of extreme physics in which particle acceleration, magnetic fields and gravitational forces far exceed those achievable in terrestrial laboratories and in which the detailed physical processes are debated. Unlike AGN jets, these stellar sources will remain spatially unresolved to all current and future technology. Only polarization measurements can directly probe magnetic fields, and constrain radiation emission mechanisms, particle acceleration and source geometry. Pioneering observations at optical and microwave energies have confirmed that early afterglow and late-time prompt emission from long GRBs are polarized (Mundell et al., 2013, Troja et al., 2017, Laskar et al. 2019). However, bright optical flashes are rare; the reason remains unclear, but is likely due to magnetization properties of the flow before and after the reverse shock that is produced when the relativistic flow first impacts the surrounding circum-burst medium – the last point at which the properties of any intrinsic magnetic field in the jet is still encoded in the detected light.  A mission like NGXP would provide the first ever measurements of GRB X-ray polarization - a unique direct probe of the magnetized outflow, at higher energy density and - for the first time – a statistically significant sample. Such measurements will revolutionize GRB theories, distinguishing between synchrotron radiation and inverse Compton processes, hadronic or leptonic plasmas, baryonic or Poynting-flux-dominated jets across a diversity of GRB central engine life-cycles.

## 2.5 Active Galactic Nuclei

Active galactic nuclei (AGNs) are the best targets to study the formation, evolution and fate of supermassive black holes and powerful persistent jets. Thanks to their high non-transient X-ray luminosities ($10^{42} - 10^{45}$ erg s$^{-1}$ up to redshift ~ 3), they are among the most efficient sources to probe both the nearby and distant Universe. More than 400 AGNs per degree squared, including radio-quiet (non-jetted) and radio-loud (jetted) objects, can be routinely detected with current X-ray observatories in the 2 – 8 keV band (Moretti et al., 2002). Yet, as of today, none of them has ever been observed in X-ray polarization despite the incredible progresses ultraviolet, optical, infrared and radio polarimetry has achieved in this field (Antonucci 1993, Urry & Padovani 1995). With the advent of IXPE and of eXTP, a handful of nearby and very bright AGNs will be observed by 2030 but the spectral resolution of the polarimetric data will be very sparse compared to the richness of the X-ray (polarimetric) signatures in the 0.1 – 100 keV band. Much of the information about the central engine (the supermassive black hole, its accretion disk and the X-ray corona) and about the parsec-scale components (circumnuclear gas/dust reservoir, ionized winds, and various distant obscurers) will remain invisible until we develop an efficient, high-resolution, X-ray polarization-focused mission. AGN observations are particularly advantageous in X-rays since the high energy band offers high contrast between the polarized AGN and the unpolarized stellar/plasma light. In addition, the penetrating power of X-rays, in particular above 5 keV, allows radiation to pass through column densities of $3 \times 10^{23}$ cm$^{-2}$ (corresponding to $A_V$ ~ 150 mags). This allows us to probe even the most dust/gas embedded objects that, maybe, are at the first stages of AGN evolution. From the central engine to the kilo-parsec scale, the variety of high energy phenomena we aim at probing is behemothic.

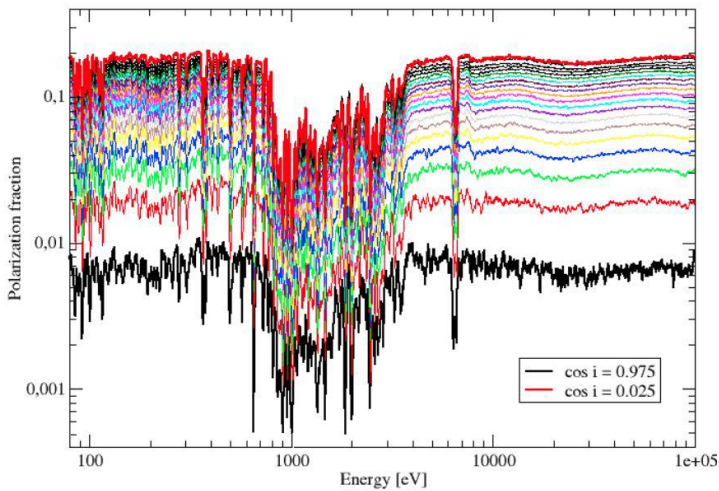

Fig. 2.4: Polarization in the 0.8 – 100 keV energy band from an ionized, multi-layered accretion disk around a supermassive black hole. Relativistic effects are not included. One can see the effect of polarization dilution by line emission (Marin et al, in prep.).

Starting from the potential well and its surrounding, high spectral resolution X-ray polarimetry from the 0.1 to the 100 keV energy will allow us to follow the complete cycle of sources's X-ray activity. Ultraviolet photons thermally emitted by the optically thick accretion disk are thought to be reprocessed by a plasma-filled region that is situated above the disk but at an unknown location and with a poorly constrained geometry. This X-ray corona will be more constrained thanks to IXPE and eXTP, but the energy band in which we will observe this essential component for radiation production in AGNs will only be probed between 2 and 8 keV, with only three or four spectral bins at best. It will not be possible to detect any line emission (if any) from the corona, nor any subtle effects such as polarized flux variations in moderate time scales (a few hours) unless a much larger X-ray polarimetric satellite is launched. By looking at the 0.1 to 10 keV band, polarimetric data will give us precious information on the corona (Poutanen & Svensson 1996, Beheshtipour et al., 2017, Tamborra et al. 2018). The same band will also allow us to detect polarization signatures from the accretion disk that is irradiated by the X-ray corona (see Fig. 2.4). Soft photons will be absorbed by the disk gas and re-emitted through X-ray fluorescent emission that, in turns, will be modified by strong Doppler and gravitational effects. A large-scale X-ray polarimetric mission will thus allow us to probe the accretion disk but also the mass and angular momentum (spin) of the supermassive black hole itself (Dovčiak et al., 2008; Schnittman & Krolik 2009, 2010). By doing so, we could independently evaluate the cumulative mass density of supermassive black holes and compare the mean accretion rate and star formation history of galaxies as a function of redshift. The higher energy band (up to 100 keV) could then tell us about the reflectivity of the disk and of the circumnuclear matter and thus its albedo-related composition. This band is particularly important since it is expected to show the most prominent polarization degree thanks to multiple Compton scattering events (Matt 1993, Poutanen et al. 1996).

In the case of radio-loud AGNs, X-ray photons may not solely come from the X-ray corona but also from the jet itself. This is particularly interesting since the production of X-ray photons in relativistic, ballistic flows of plasma is not entirely understood. There are usually three scenarios used to explain this emission: Compton scattering of blackbody photons emitted from the accretion disk, scattering of Broad Line Region photons and self-Comptonization of intrinsically polarized synchrotron photons emitted by jet electrons (scattering of CMB photons may be relevant in the outer parts of the jets). It was shown (Celotti & Matt 1993, Poutanen 1994) that synchrotron self-Comptonization is distinguishably with respect to the other two mechanisms in terms of both polarization degree and angle. SED and polarization properties from SSC in a multizone jet scenario are shown in Fig. 2.5 (Peirson & Romani 2019). One may also see highly polarized X-ray emission for a proton synchrotron component, if present (Zhang & Boettcher 2013). A leptonic jet is expected to show energy-dependent signatures in the 0.1 – 100 keV regime that is quite distinctive from a hadronic model, but only a large coverage of the energy band can detect the variations (see Figure 2.6). With an improved X-ray polarimetric imager that goes beyond 10 keV, structures in the jet of nearby radio-loud AGNs could be distinguished similarly to what has been observed in the radio band (Martí-Vidal et al., 2012). Then, correlations between the radio and X-ray band emission and polarization would be immediate and tell us a lot about jet structure and magnetic field topology.

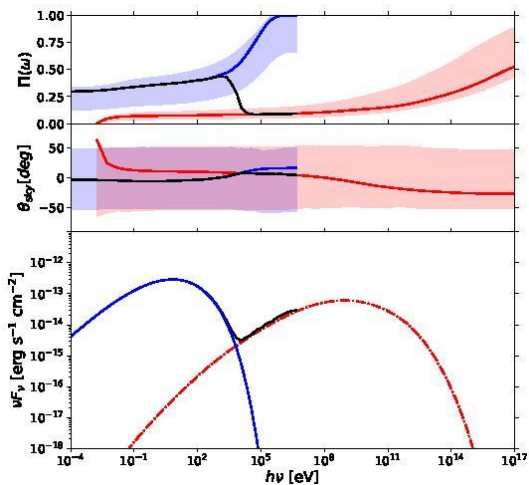

Fig. 2.5: SED and polarization behavior from a multi-zone High Synchrotron Peak jet with Lorentz Factor $\Gamma=14$ and electron power law index $\alpha=1.85$ viewed at $\vartheta=6°$ showing synchrotron (blue) and SSC (red) emission. The polarization and EVPA range across the shaded bands in individual realizations; a typical realization is show by the curves. Note the rise in polarization degree $\pi$ toward the end of each SED component (hard X-ray for synchrotron) and the EVPA step as one moves from the synchrotron to Compton regimes in the X-ray band (after Peirson & Romani 2019)

With higher sensitivity X-ray polarimetric satellites, it should be possible to have a look at the other AGN components. In particular, highly ionized, ultra-fast outflowing winds can tell us a lot about accretion-ejection processes around compact objects (Tombesi et al., 2010, 2011, 2012). Observing the wind features in polarimetric data could lead to a better identification of the wind composition and geometry. The degree and position angle of polarization is significantly altered by the morphological parameters of the wind (launching radius, opening angle, collimation angle if the wind's interior is hollow, see Marin & Goosmann 2013). Such constraints could then be fed to magneto-hydrodynamical models to produce more realistic winds and thus better estimate the amount of energy and matter that is deposited in the close environment of AGNs (Proga et al., 2007). X-ray polarimetry could then independently constrain the feedback processes of AGN that may lead to star formation, quenching or re-activation, a phenomenon yet to be observed. A morphological determination of the ejection winds will also lead to a better determination of the luminosity-dependent size of the circumnuclear gas/dust region that is surrounding the central engine along the equatorial plane. This structure, essential for the comprehension of AGNs, is yet to be resolved in X-rays and its physical limits and opening angles are not known. With X-ray polarization, it would be possible to determine the height of the structure but also its hydrogen column density. The amount of gas will directly impact the energy at which X-ray radiation will be able to penetrate this region and, for an equatorial observer, this will result in a rotation of the polarization position angle that can be used as a very precise probe for the gas. Since this perpendicular rotation of the polarization angle is case-sensitive, it is also energy-dependent and only a wide coverage of the X-ray band can be used to estimate the hydrogen column density without fail (Marin et al., 2018a, 2018b).

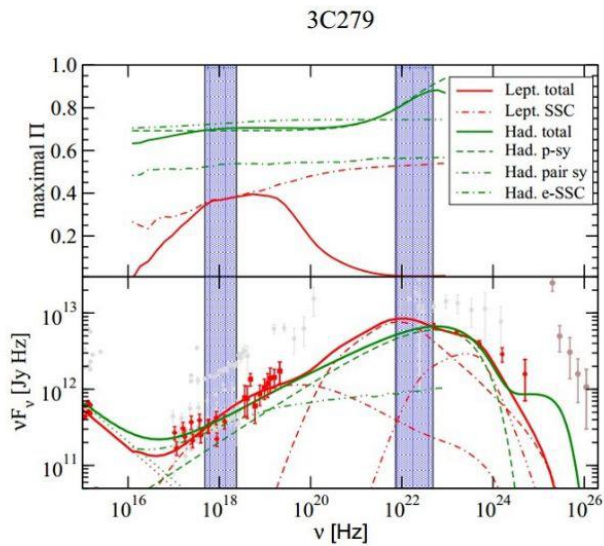

Fig. 2.6: UV through γ-ray SEDs (lower panels) and the corresponding maximum degree of polarization (upper panels) for 3C279. Leptonic model fits are plotted in red, hadronic models in green. Different line styles indicate individual radiation components, as labelled in the legend (Zhang & Boettcher, 2013).

## 2.6 SGR A* and the Galactic Center

Detection of hard X-ray emission (Markevitch et al., 1993; Sunyaev et al., 1993) and the iron fluorescent line at 6.4 keV (Koyama et al., 1996) co-spatial with the most prominent molecular complexes in the Galactic center region gave rise to a hypothesis that that we are witnessing the light-travel-time-delayed X-ray 'echo' of a powerful flare produced by the supermassive black hole Sgr A* some hundreds of years ago. This hypothesis got strong further support from the observed variability of the reflected emission (see Ponti et al., 2015, for a review), which is consistent with the theoretical expectations (Sunyaev, & Churazov, 1998). In addition to reconstruction of the Sgr A*'s past activity record (e.g. Clavel et al., 2013; Chuard et al., 2018), one could use the same data as a sensitive and bias-free diagnostic tool for the illuminated gas (Molaro et al., 2016; Churazov et al., 2017; Khabibullin et al., 2019). Thanks to the X-ray reflection, eventually we will be able to map the 3D density (and potentially velocity) field of the dense gas inside inner a few hundred pc of our Galaxy (Marin et al., 2014, 2015; Churazov et al., 2017) and possibly some other nearby galaxies (e.g. Clavel et al., 2013; Fabbiano et al., 2019; Churazov et al., 2019).

Besides hard spectral shape and temporal variability, this paradigm predicts high level of polarization for the observed X-ray continuum, with the exact value being sensitive to the relative line-of-sight position of the reflecting cloud (Churazov et al., 2002). In the simplest scenario of a single flare, the degree of polarization in continuum is set by a the cosine of scattering angle, $\mu$, so that $P = (1-\mu^2)/(1+\mu^2)$, reaching 100% for 90-degrees scattering. At the same time, direction of the polarization plane should be perpendicular to the line connecting the cloud and the primary illuminating source. Thanks to this, measurements of the X-ray polarization would provide a dramatic boost to the diagnostic power of both the small-scale (i.e. intra-cloud) and the large-scale (i.e. across the Central Molecular Zone) probes. In particular, i) the direction of the polarization plane allows us to constrain localization of the primary source, potentially proving that it is indeed Sgr A*, ii) the degree of polarization gives the line-of-sight position of the scattering material, and iii) polarization data facilitates cleaner separation of the scattered emission from various unpolarized backgrounds and foregrounds. The predicted map of the polarization degree for the reflected X-ray emission is shown in Figure 2.6, as calculated by Churazov et al., (2017) based on the gas density distribution model for the Central Molecular Zone of the Galaxy by Kruijssen et al., (2015) (see also Marin et al., 2015, for a similar simulation). A Sgr B2-based example of the expected appearance of the polarization degree as a function of energy for several scattering angles is illustrated in Fig. 2.7, where neither unpolarized background or foreground is removed from the spectrum.

Some of these diagnostics will become feasible already with the forthcoming X-ray polarization missions like IXPE (Weisskopf et al., 2016), while the future X-ray missions with higher sensitivity and better angular resolution will open new exciting possibilities (e.g. Churazov et al., 2019). Once Sgr A is confirmed as a primary source and the parameters of the flare (duration and the time elapsed since the onset of the outburst) are well established, it will become possible to constrain the intrinsic polarization of the primary Sgr A*'s emission as a function of viewing angle, and potentially shed light on the physical mechanism behind it. This could, in turn, be used to explain the X-ray chimney-like structures detected in the Galactic Centre (Nakashima et al., 2013; Ponti et al., 2019). Those chimneys, filled with hot plasma, might very well be the physical and energetic link between the central galactic region and the large-scale Fermi bubbles (Su et al., 2010).

A further step forward would be to extend similar diagnostics to central regions of other galaxies (Fabbiano et al., 2019) and study the "past" flares from currently quiescent AGNs and asymmetries in the gas distribution around them. In this regard, there will be a very interesting possibility to use the sample of a thousand X-ray-bright Tidal Disruption Events to be detected during the course of the SRG/eROSITA All-Sky Survey (Khabibullin et al., 2014). The main parameters of these flares will be well constrained, so one can reconstruct the location of the illuminated gas at any given moment. For the 2050 time frame, it implies that 5 – 30 pc region around SMBHs can be systematically studied.

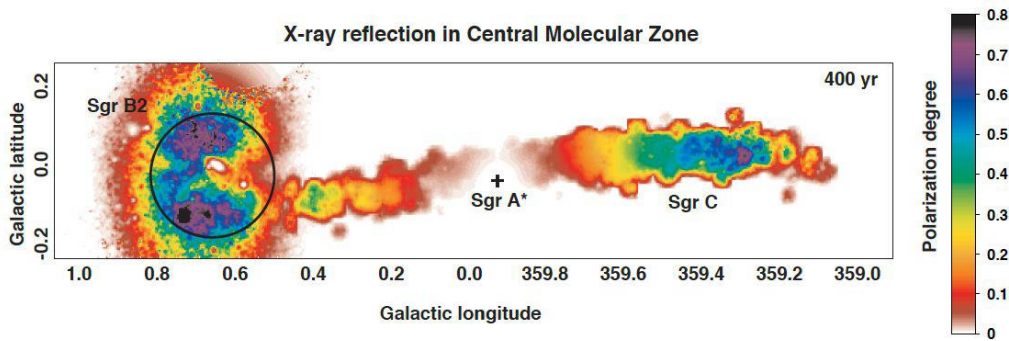

Fig.2.6: Expected degree of polarization for the continuum in the 4 – 8 keV band (excluding fluorescent lines) adApted from Churazov et al. (2017). The colorbar to the right shows the degree of polarization. The 50 yr long outburst model is used to simulate the reflected emission 400 yr after the onset of the outburst. For the very central regions, the degree of polarization is low, since the scattering angle is close to 180°. The largest degree of polarization (tens of percents) is expected for the Sgr B2 cloud, since in these simulations it was assumed to be located at the same line-of-sight distance as Sgr A*. From the existing Chandra and XMM-Newton observations, we already know that the matter distribution is more complicated. The degree of polarization is effectively a proxy to the location of the scattering cloud along the line-of-sight.

Finally, X-ray reflection of the central SMBH's emission is, of course, not limited to the inner regions of the galaxies. As has been pointed out long ago (Vainshtein & Syunyaev, 1980), it allows in complementary studies of the galactic ISM at kpc scales and SMBH's activity record over thousands of years (Cramphorn & Sunyaev, 2002). Clearly, X-ray polarization will be a very efficient way to search for such signals on top of the collective X-ray emission of faint unresolved galactic sources (e.g. the Galactic Ridge emission, Revnivtsev et al., 2009) and also the reflected emission of the brightest (possibly transient) sources (Molaro et al., 2014). The latter point is particularly relevant for the so-called Ultra-Luminous X-ray Sources, which have apparently super-Eddington X-ray luminosity but could be characterized by high degree of collimation. Searching for the reflection of their emission on the surrounding ISM (including molecular clouds) might help constraining their angle-averaged luminosity and also allows finding 'misaligned' ULXs (akin the Galactic super-accretor SS 433, Khabibullin, & Sazonov, 2016, 2019).

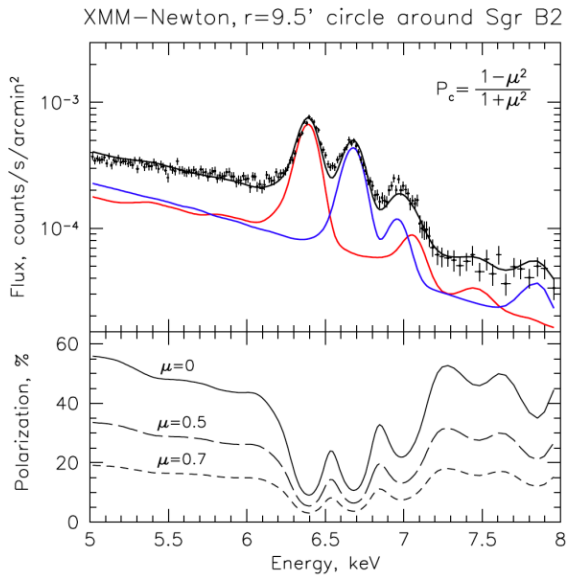

Fig.2.7: Top: spectrum extracted from a circle around Sgr B2 clouds (XMM-Newton data, see Churazov et al., (2017). The red and blue lines show the decomposition of the spectrum into the reflection and hot plasma components. Bottom: expected degree of polarization of the pure reflection component as a function of energy for different scattering angles set by positions of the molecular clouds with respect to Sgr A*.

## A CONCEPT FOR A NEXT GENERATION X-RAY POLARIMETRY MISSION (NGXP)

With the launch of IXPE, scheduled on April 2021, X-ray Polarimetry will pass the infancy phase. IXPE was conceived to make a first assessment of the polarization properties of a number of classes of X-ray emitters in the frame of a cheap and fast mission. It is a matter of fact that the two polarimetric missions included in the final selection of the SMEX call (IXPE and PRAXyS) were active in the classical 2 – 8 keV X-ray band. Moreover, the mission studied by ESA for the M4 slot (XIPE) was also sensitive in this band, so that after the selection of IXPE by NASA, ESA evaluated that XIPE would extend the sample but not add any other improvement except in sensitivity. It is a fact that, in the total absence of data, the classical band is the one for which a reasonable number of predictions were available in literature and a reasonable efficiency, compliant with a good deal of these predictions, could be achieved in the frame of a small mission. Moreover, this band maximizes the number of classes of X-ray sources which can be studied, and therefore the fraction of the X-ray community to be involved. It is therefore not surprising that the teams of IXPE, PRAXyS and XIPE, facing also mass and budget limitations, decided to restrict themselves to the classical band. Assuming that IXPE will be successful in a good fraction of its targets, and data will give a more quantitative assessment of the capabilities of this technique, the pressure from the community for a much more sensitive instrument, capable to improve significantly the measurements of IXPE and to extend to other topics that IXPE could not attack, will justify a much more effective effort for a mission of the medium size class.

While waiting for IXPE data for a more precise assessment of the polarization levels and observation requirements, some points are already well established.
1) Polarimetry is a photon-hungry technique. IXPE is a small mission of 300 kg. An increase of an order of magnitude of the effective area is required for the next step.
2) The band should be extended to lower energies (0.1 – 0.2 keV). The use of beryllium windows, and the intrinsic properties of photoelectric effect, have excluded some important objectives in the study of isolated Neutron Stars, blazars, BH binaries, Tidal Disruption Events (TDE).
3) The band should be extended to higher energies (up to 50 – 70 keV) to include synchrotron lines, non-thermal components of shocked plasmas in SNR, the hardening of spectrum of magnetars. Last but not least a good continuity in the response from 5 to 50 keV should be guaranteed to seriously attack models involving reflection, of paramount importance in all high energy astrophysics.

4) The angular resolution is important in general but very important for some specific topics such as turbulence in SNR or PWN mapping. An improvement of, at least, a factor 5 is hoped.
5) Most of X-ray sources are highly variable in flux and spectrum and likely in polarization too. For most classes the different behaviour has been classified in terms of different status. An efficient polarimetric observing plan would enormously benefit of a previous knowledge of the status of sources or of the existence of newly discovered sources.
6) Some bright sources come from unpredictable directions in the sky, and to catch them requires a very large field of view. Obvious examples are GRBs and TDE.

These are the major improvements to be achieved with a new design. All of them advocate for a mission of the medium size class. A design - even a simple one - is not needed in this phase. We try to identify a concept by using as building blocks studies performed to various states of advancement. We will base our proposal on well-established concepts and on studies already performed in a recent past and publicly available, and we will only mention areas of improvement, based on facts as well.

***The available techniques.*** Techniques available to the community can be grouped according to the basic physical process:
a) Bragg Diffraction tuned at low (2 – 8 keV) or very low (0.2 – 2 keV) energies depending on the diffracting material. Due to the very narrow band, above 2 keV, cannot compete with photoelectric devices.
b) Photoelectric polarimetry in gas with Gas Pixel Detector (Imaging) or with Time Projection Chamber (not imaging). Effective at low energies (2 – 8 keV) with low Z gas filling (Ne or DME) and at medium energies (6-20 keV) with A filling.
c) Scattering polarimetry. Devices are called active when both scatterer and absorber are detectors in coincidence, and passive when the absorber is the only detector. The materials of the two stages can be different (with the scatterer of lower atomic number) or the same.

***A multi-telescope system.*** The combination of guidelines listed above cannot be efficiently fulfilled with a single telescope design. The extension to higher energies set a ratio between the diameter and the length. A single telescope would require an amazingly long focal or a formation flight. Incidentally, we stress the point that while managing the problem of stray light or sky light with a deployable system is tough, with a formation flight it is almost impossible. In some cases, techniques to perform different measurements can be stacked or combined in the same telescope. In some case, however, the combination would be very inefficient. Two detector units would be the optimal solution.

Points 1) and 4) and the just mentioned arguments point straightforwardly toward a multi-mirror assembly with a focal length of the order of 10m. This could be achieved with one or more deployable devices as actuated in NUSTAR (Harrison et al., 2013), Hitomi (Takahashi et al., 2018) and IXPE (Weisskopf et al., 2016) and foreseen for NHXM (Tagliaferri et al., 2010). Yet we do not propose such solution which is more typical of a small mission carried by a small launcher (such as Pegasus). In a multi-mirror system it is highly desirable a certain rigidity and repeatability of the positions in the focal plane to ensure the relative calibration and stability in control of systematics. Launchers with a fairing able to accomodate telescopes 10 m long are nowadays available. Following the example of M4 mission ARIEL we assume that a Medium Size European mission could be launched by an Ariane-6 rocket and operated in L2. This would expand the observing time, remove the orbital gaps and make easier the interaction with the ground segment, adding flexibility to the observing program. In alternative a high eccentricity orbit (e.g. EXOSAT, Chandra, Newton, INTEGRAL) is viable giving continuous observations of days.

***Classical Energy range Imaging Polarimeter.*** Imaging polarimetry is vital for various classes of sources and the NGXP Mission should improve the angular resolution. The imaging Polarimeter is based on the Gas Pixel Detector. The imaging capabilities of this device are limited. While the margins to improve the quality of the optics are wide, the detector has some intrinsic limitation due to the finite space resolution and the effect of the finite thickness in presence of inclined beam arriving from the optics. Given that the inclination is defined by the energy band, we can state that in the classical range 2 – 8 keV a half power radius of 250 µm is a minimum. A realistic goal is 10" HPR, with 5" as a more ambitious one. With present technology, optics compliant with these goals can be achieved. E.g. Silicon optics are a viable technique. While we assume that after IXPE any mission of X-ray polarimetry should include an imaging capability, we also assume that this quality can be limited to two of the telescopes whose collecting area can represent a significant improvement beyond IXPE. We adopt the design of XMM that fits the selected envelope. We only assume an increase of area thanks to the 30 years of technical developments.

*Hard X-ray Imaging Polarimeter.* The traditional application of photoelectric polarimetry is based on low Z gas mixtures (Ne or DME). The technique could be applied to higher energies with higher atomic number gases, but with decreasing efficiency because of the high Auger electron energy and the less favourable stopping/scattering ratio. But it has been already demonstrated that the technique is viable with Argon based mixtures with an increased absorption gap thickness and with gas pressure of the order of 3 atmospheres. Such a detector has an imaging capability about twice worse than the DME GPD and can extend to the 6–25 keV band the capability to make polarimetry totally source dominated. The sensitivity in terms of MDP is comparable to that of a DME GPD for a comparable milli-crab flux. These detectors have the same problem of other GPDs, namely they cannot be transparent to photons which cross the gas. Imaging detectors in this energy band would be useful to measure angularly resolved polarization of non-thermal emission from Pulsar Wind Nebulae and Supernova Remnants. Also the sensitivity to Molecular Clouds around the Galactic Center region would be increased due to reflection nature of their emission. Moreover, the hot plasma emission component would be minimized.

We propose to include two imaging telescopes based on this technology.

*Wide Range Composite Polarimeters (non-Imaging).* Most X-ray sources are point-like. The need for imaging is limited to those in crowded fields or so weak to be comparable to the background inside the field of view. The study of point like sources can be performed with non-imaging focal plane polarimeters, provided they have a field of view narrow enough to avoid source confusion, but an adequate polarimetric sensitivity. Thus we adopt a non-imaging configuration for some
of the telescopes, to allow a very broad-band polarization sensitivity, including<keV and >25keV, where true imaging polarimetry is difficult or impossible. The photoelectric polarimeters based on GPD and those based on TPC are similar in terms of sensitivity when the imaging is not required. The correct use of TPC polarimeter requires however rotation. The GPD collects the charges from the side opposite to the window and, therefore, photons crossing unperturbed the gas, do not pass the silicon chip, the ceramic case and the supporting PCB. In the TPC the electrons are drifted and collected on the side. The bottom of the detector can be as transparent as the front window. By a proper tuning of the gas pressure and the gap thickness the TPC can absorb photons of selected lower energy and be transparent to photons of higher energies.

In lower energies band the Bragg diffraction is not very effective because of the very narrow band. By combining three diffraction gratings and three graded multilayer organic diffractor (artificial Bragg) a focal plane polarimeter, effective around 0.2 keV, can be made with a wider energy band compared with the 'classical' use of diffractors.

At higher energies, telescopes with multilayer coating (like NUSTAR) can focus photons up to the energy of the K edge of the coating material. At energies where the photoelectric effect is too inefficient, focal plane polarimetry can be performed with a stick of low atomic number material (lithium, or Beryllium or Plastic Scintillator) in the focus surrounded by a well of detectors to detect the azimuth modulation of scattered photons. The scatterer can be active (plastic scintillator in coincidence with the absorbers) or passive (lithium or beryllium). The active approach is limited to higher energies. The passive configuration is limited by higher background.

A combination of the three techniques in the focus of a single telescope has been studied also in the frame of the XPP proposal (Krawczynski et al., 2019b, Jahoda et al., 2019). The same optics is used to make broad band polarimetry. This approach can be a good baseline. Yet the presence of a Hard X-Ray Imaging polarimetry could cover well the 6 – 20 keV range, while leaving to an active scattering polarimeter in the stacked configuration.

*Wide Field Camera.* The transient nature of the X-ray sky, the capability to observe X-ray sources in different states with different expected polarization, the possibility to repoint towards previously unknown sources including GRBs requires an on-board All Sky Monitor (ASM). Two different kind of ASMs are currently available. One is based on the use of the so-called lobster eye technology. This true 2-D technology is effective only below 2 keV leaving unknown the nature of many new found sources. The other makes use of a coded mask coupled with 1-D silicon detectors as the Silicon Drift Detectors. This technology is at the base of the design of missions like LOFT or eXTP. The field of view is effective between 2-50 keV, the proper energy range to distinguish between states and kind of unknown sources. It requires a couple of sensors arranged at 90 deg for X and Y detection. There will be 4 couples of them around the pointing direction of the telescopes. The whole field of view will be one-steradian at zero-response with a location accuracy of 1 arcmin and an angular resolution of 5 arcmin. The sensitivity is 3 mCrab at 5-sigma in 50 ksec. The energy resolution of 200 eV will be approximately constant in the energy range of this instrument. We expect with this configuration 1 GRB detection every four days. With this technique a source localisation well below one arcminute is possible. This allows to identify known sources and derive their status to trigger re-pointing. It also allows to subtract the false modulation due to off axis source of the GRB polarimeter described below. But this is especially suited to trigger TOO re-pointing of transients and first of all of GRBs.

*Gamma-Ray Burst Polarimeter.* The science of GRBs has made important progresses since the discovery of extragalactic location. Data from Swift have provided, thanks to a rich phenomenology of the afterglow X-ray emission, a first way to classify these elusive phenomena. Theoretical analysis has shown that polarimetry could be a powerful tool to discriminate between the different models which are consistent with GRB's spectra and timing. Unfortunately, polarimetry of prompt GRBs is committed to small instruments or to instruments designed for other purposes, giving polarimetry as a by-product. X-ray Polarimetry of afterglows is totally absent. We want to perform polarimetry of both prompt and afterglow of the same burst for a sample of at least 20 GRBs per year. Polarimetry of prompt emission has been studied by many groups, based on Compton scattering detected from an array of solid detectors. There are strong arguments in favour of a combination of low atomic number scintillators, acting as scatterer and high atomic number scintillators acting as absorbers. The use of fast inorganic scintillators (as GAGG, Pearce et al. 2019, Kushwah et al. 2019) and Silicon Photomultipliers, expands the capability of this approach. Two such detectors will cover a 1.5 $\pi$ field including the direction of telescopes. Beside performing the polarimetry of the prompt, the mission will be able to detect GRBs with WFC, localise them in a few seconds and steer the pointing of the telescopes according to the Swift approach.

Beside the paramount relevance of GRBs, polarimetry of other transients could be very fruitful. In general polarimetry of any time tagged object can be done. This includes accretion and rotation powered pulsars.

| Telescopes | Imaging | # | 0.2 | 2-8 | 6-25 | 15-80 |
|---|---|---|---|---|---|---|
| Classical Range Energy Polarimeter | Y | 2 | | X | | |
| Hard X-ray Energy Polarimeter | Y | 2 | | | X | |
| Broad Band Polarimeter | N | 4 | X | X | X | X |
| **Wide Field Instruments** | | | | | | |
| Wide Field Camera | Y | 4 | | | | |
| Transient Sources Polarimeter | N | 2 | | | | |

# References

Aharonian, F. et al., 2018, PASJ, 70, 113
Antonucci, R., 1993, ARA&A, 31, 473
Beheshtipour, R., Krawczynski, H., Malzac, J., 2017, ApJ, 850, 14
Bykov, A. M., Uvarov, Yu. A., Bloemen, J. B. G. M., den Herder, J. W., Kaastra, J. S., 2009, MNRAS, 399,1119B
Bykov, A. M., Churazov, E. M., Ferrari, C., et al., 2015, Space Sci Rev, 188, 141
Bykov, A., Uvarov, Y., 2017, J Phys: Conf. Ser., 932, 012051
Campana, S., Di Salvo, T., 2018, arXiv:1804.03422
Chaty, S., Mirabel, I. F., Duc, P. A., Wink, J. E., Rodriguez, L.F., 1996, A&A, 310, 825
Cheng, Y., Liu, D., Nampalliwar, S., Bambi, C., 2016, Classical and Quantum Gravity, 33, 125015
Chuard, D., Terrier, R., Goldwurm, A., et al., 2018, A&A, 610, A34
Churazov, E., Sunyaev, R., & Sazonov, S., 2002, MNRAS, 330, 817
Churazov, E., Zhuravleva, I., Sazonov, S., et al., 2010, Space Sci Rev, 157, 193
Churazov, E., Khabibullin, I., Sunyaev, R., et al., 2017, MNRAS, 471, 3293
Churazov, E., Khabibullin, I., Ponti, G., et al., 2017, MNRAS, 468, 165
Churazov, E., Khabibullin, I., Sunyaev, R., et al., 2017, MNRAS, 465, 45
Churazov, E., Khabibullin, I., Sunyaev, R., et al., 2019, Bulletin of AAS, 51, 325
Celotti, A., Matt, G. 1994, MNRAS, 268, 451
Chauvin, M., et al., 2017, Nature Scientific Reports, 7, 7816
Chauvin, M., et al., 2018a, MNRAS, 477, L45
Chauvin, M., et al., 2018b, Nature Astronomy, 2, 652
Chauvin, M., et al., 2019, MNRAS, 483, L138
Clavel, M., Terrier, R., Goldwurm, A., et al., 2013, A&A, 558, A32
Clavel, M., Terrier, R., Goldwurm, A., et al., 2017, The Multi-messenger Astrophysics of the Galactic Centre, 253
Cramphorn, C. K., Sunyaev, R. A., 2002, A&A, 389, 252
Cramphorn, C. K., Sazonov, S. Y., Sunyaev, R. A., 2004, A&A, 420, 33
Connors, P. A., Stark, R. F., Piran, T., 1980, ApJ, 235, 224
Costa E., et al., 2001, Nature, 411, 662
de Ona Wilhelmi, E., et al., XIPE Science Working Gro 2017 AIP Conf. series 1792, 070023
Dermer, C. D., Schlickeiser, R., Mastichiadis, A., 1992, A&A, 256, L27
Di Matteo T., Blackman E.G., Fabian A.C., 1997, MNRAS, 291, L23
Dodson, R., et al., 2003, MNRAS, 343, 116
Dovčiak, M., Muleri, F., Goosmann, R. W., Karas, V., Matt, G., 2008, MNRAS, 391, 32
Dovčiak, M., Muleri, F., Goosmann, R. W., Karas, V., Matt, G., 2011, ApJ, 731, 75
Fabbiano, G., Siemiginowska, A., Paggi, A., et al., 2019, ApJ, 870, 69
Fabian A. C., et al., 2017, MNRAS, 467, 2566
Fender, R., 2001, MNRAS, Volume 322, 31
Forot M., et al., 2008, ApJ, 688, L29
Fortin, J. F, Kuver, S., 2019, JHEP, 2019, 163
Fragile, P. C., Blaes, O. M., Anninos, P., Salmonson, J. D., 2007, ApJ, 668, 417
Gaensler, B. M., Slane, P. O., 2006, ARAA, 44, 17
Gilfanov, M. R., Syunyaev, R. A., Churazov, E. M., 1987, Soviet Astronomy Letters, 13, 233
González Caniulef, D., Zane, S., Taverna, R., Turolla, R., Wu, K., 2016, MNRAS, 459, 3585
Gotz D., et al., 2014, MNRAS, 444, 2776

Harrison F., et al., 2013, ApJ, 770, 103
Helder, E. A., Vink, J., Bykov, A. M., Ohira, Y., Raymond, J. C., Terrier, R., 2012, SSRv, 173, 369H
Heyl, J. S., Shaviv, N. J., 2000, MNRAS, 311, 555
Heyl, J. S., Shaviv, N. J., 2002, Phys. Rev. D, 66, 023002
Hua, X.-M., Titarchuk, L., 1995, ApJ, 449, 188
Ingram, A., Done, C., Fragile, P. C., 2009, MNRAS, 397, 101
Ingram, A., Maccarone, T. J., Poutanen, J., & Krawczynski, H., 2015, ApJ, 807, 53
Israel G. L., et al., 2008, ApJ, 685, 1114
Jahoda, K., et al., 2019, arXiv:1907.10190
Kargaltsev, O., Pavlov, G. G., 2008, AIP Conference Proceedings, 983, 171
Kaspi, V. M., Beloborodov A. M., 2017, ARA&A, 55, 26
Kaspi, V. M., 2010, PNAS, 107, 7147
Khabibullin, I., Sazonov, S., 2016, MNRAS, 457, 3963
Khabibullin, I., Churazov, E., Sunyaev, R., et al., 2019, arXiv:1906.11579
Khabibullin, I., Sazonov, S., Sunyaev, R., 2014, MNRAS, 437, 327
Khabibullin, I. I., Sazonov, S. Y., 2019, Astronomy Letters, 45, 282
Kislat, F., et al., 2018, JATIS, 4, 011004.
Kothes, R., et al., 2008, ApJ, 687, 516
Komarov, S. V., Khabibullin, I. I., Churazov, E. M., et al., 2016, MNRAS, 461, 2162
Koyama, K., Petre, R., Gotthelf, E. V., Hwang, U., Matsuura, M., Ozaki, M., Holt, S. S., 1995, Nature, 378, 255
Koyama, K., Maeda, Y., Sonobe, T., et al., 1996, PASJ, 48, 249
Krawczynski H., et al., 2019b, arXiv:1904.09313
Kruijssen, J. M. D., Dale, J. E., Longmore, S. N., 2015, MNRAS, 447, 1059
Kushwah R., et al. 2019, NIM A, in press
Kylafis N.D., Belloni T.M., 2015, A&A, 574, 133
Lapidus, I. I., Sunyaev, R.A., 1985, MNRAS, 217, 291
Laskar et al. 2019, ApJL, 878, 26
Laurent P., et al., 2011, Science, 322, 338
Machine et al 2007, JCAP, 10, 13
Maraschi, L., Ghisellini, G., Celotti, A., 1992, ApJ, 397, L5
Marin, F., Goosmann, R.W., 2013, MNRAS, 436, 2522
Marin, F., Karas, V., Kunneriath, D., Muleri, F., 2014, MNRAS, 441, 3170
Marin, F., Muleri, F., Soffitta, P., et al., 2015, A&A, 576, A19
Marin, F., et al., 2018a, MNRAS, 473, 1286
Marin, F., Dovciak M., Kammoun, E., 2018b, MNRAS, 478, 950
Markevitch, M., Sunyaev, R. A., Pavlinsky, M., 1993, Nature, 364, 40
Marti-Vidal, I., et al., 2012, A&A, 542, 107
Matt, G., 1993, MNRAS, 260, 663
McNamara, A. L., Kuncic, Z., Wu, K., 2009, MNRAS, 395, 1507
Mignani, R.P., et al., 2017, MNRAS, 465, 492
Molaro, M., Khatri, R., Sunyaev, R. A., 2014, A&A, 564, A107
Molaro, M., Khatri, R., Sunyaev, R. A., 2016, A&A, 589, A88
Moretti A., et al., 2002, ApJ, 570, 502
Mundell et al. 2013, Nature, 504, 119
Nakamura, Y., Shibata, S., 2007, MNRAS, 381, 1489

Nakashima, S., Nobukawa, M., Uchida, H., Tanaka, T., et al., 2013, ApJ, 773, 20
Nättilä, J., Miller, M.C., Steiner, A.W., et al., 2017, A&A, 608, A31
Ng, C.Y., Romani, R.W. 2007, ApJ, 660, 1357
Noutsos, A., Kramer, M., Carr, P., Johnston, S., 2012, MNRAS, 423, 2736
Patruno, A., Watts, A.L., 2012, arXiv:1206.2727
Pavlov G.G., Zavlin V.E., 2002, ApJ, 529, 1011
Pearce M. Et al., 2019, Astroparticle Physics, 104, 54
Peirson A.L., Romani R.W., 2019, ApJ, submitted
Ponti, G., Morris, M. R., Terrier, R., et al., 2015, MNRAS, 453, 172
Ponti, G., Hofmann, F., Churazov, E., et al., 2019, Nature, 567, 347
Porth, O., et al., 2014, MNRAS, 438, 278
Postnov, K., Shakura, N. I., Staubert, R., 2013, MNRAS, 435, 1147
Poutanen, J., 1994, ApJS, 92, 607
Poutanen, J., Vilhu, O., 1993, A&A, 275, 337
Poutanen J., Nagendra K.N., Svensson R., 1996, MNRAS, 283, 892
Poutanen, J., Svensson, R., 1996, ApJ, 470, 249
Poutanen, J., Gierlinski, M., 2003, MNRAS, 343, 1301
Poutanen, J., Nättilä, J., Kajava, J.J.E., et al., 2014, MNRAS, 442, 3777
Proga, D., 2007, ApJ, 661, 693
Ramsey, B. D., 2005, Exp. Astron, 20, 85
Remillard, R. A., McClintock, J. E., 2006, ARA&A, 44, 49
Revnivtsev, M., Sazonov, S., Churazov, E., et al., 2009, Nature, 458, 1142
Revnivtsev, M. G., Churazov, E. M., Sazonov, S. Y., et al., 2004, A&A, 425, L49
Rodríguez Castillo, G. A., et al., 2016, MNRAS, 456, 4145
Santangelo, A., et al., 2019, SCPMA, 6229, 505S
Sazonov, S. Y., Sunyaev, R. A., Cramphorn, C. K., 2002, A&A, 393, 793
Sazonov, S. Y., Churazov, E. M., Sunyaev, R. A., 2002, MNRAS, 333, 191
Schnittman, J. D., Krolik, J. H., 2009, ApJ, 701, 1175
Schnittman, J. D. Krolik J. H., 2010, ApJ, 712, 908
Shklovsky I.S., 1953, Dokl. Akad. Nauk SSSR, 90, 983
Spruit, H., Phinney, E. S., 1998, Nature, 393, 139
Su, M., Slatyer, T. R., Finkbeiner, D. P., 2010, ApJ, 724, 1044
Sunyaev, R. A., Titarchuk, L. G., 1980, A&A, 86, 121
Sunyaev, R. A., 1982, Soviet Astronomy Letters, 8, 175
Sunyaev, R. A., Titarchuk, L. G., 1985, A&A, 143, 374
Sunyaev, R. A., Markevitch, M., Pavlinsky, M., 1993, ApJ, 407, 606
Sunyaev, R., Churazov, E., 1998, MNRAS, 297, 1279
Taam, R. E., Chen, X., Swank, J. H., 1997, ApJ, 485, L83
Taam, R. E., Chen, X., Swank, J. H., 1997, ApJ, 485, L83
Takahashi T., et al., 2018, JATIS, 4, 021402
Tagliaferri G., et al., 2010, Proceedings of the SPIE, 7732, 217
Tamborra F., Matt G., Bianchi S., Dovciak M., 2018, A&A, 619, 105
Taverna, R., Muleri, F., Turolla, R., Soffitta, P., Fabiani, S., Nobili, L., 2014, MNRAS, 438, 1686
Taverna, R., Turolla, R., González Caniulef, D., Zane, S., Muleri, F., Soffitta, P., 2015, MNRAS, 454, 3254
Taverna, R., Turolla, R., 2017, MNRAS, 469, 3610

Tetarenko, B. E., Sivakoff, G. R., Heinke, C. O., Gadstone, J. C., 2016, ApJS, 222, 15
Thompson C., Duncan R. C., 1995, MNRAS, 275, 255
Tiengo A., et al., 2013, Nature, 500, 312
Tombesi, F., et al., 2010, A&A, 521, 57
Tombesi, F., et al., 2011, ApJ, 742, 44
Tombesi, F., Cappi, M., Reeves, J. N., Braito, V., 2012, MNRAS, 422, L1
Troja et al. 2017, Nature, 547, 425
Turolla, R., Zane, S., Watts, A. L., 2015, RPPh. 78, 6901
Urry, C. M., Padovani, P., 1995, PASP, 107, 803
Vainshtein, L. A., Syunyaev, R. A., 1980, Soviet Astronomy Letters, 6, 353
Viironen, K., Poutanen, J., 2004, A&A, 426, 985
Vink, J, Zhou, P., 2018, Galaxies 6, 46
Volpi, D., et al., 2008, A&A, 485, 337
Wadawale S.V., et al., 2017, Nature Astronomy, 34, 1
Weisskopf, M. C., et al., 1978, ApJ, 220, L117
Weisskopf, M. C., et al., 2016, Proc. SPIE, 9905, 990517
Zhang, H., Boettcher, M., 2013, ApJ, 774, 18
Zhang, S. N., et al., 1997, ApJ, 479, 381
Zhang, S. N., et al., 2019, SCPMA, 62, 029502
Zhuravleva, I. V., Churazov, E. M., Sazonov, S. Y., et al., 2010, MNRAS, 403, 129

# Proposing Team

*Niccolo' Bucciantini (INAF, Oss. Arcetri, Italy)*
*Eugene Churazov (MPA, Germany; IKI, Russia)*
*Enrico Costa (INAF, IAPS, Italy)*
*Michal Dovciak (Academy of Sciences of the Czech Rep.)*
*Hua Feng (Tsinghua Univ., China)*
*Jeremy Heyl (Univ. British Columbia, Canada)*
*Adam Ingram (Univ. of Oxford, U.K.)*
*Keith Jahoda (NASA/GSFC, U.S.A.)*
*Philip Kaaret (Univ. of Iowa, U.S.A.)*
*Timothy Kallman (NASA/GSFC, U.S.A.)*
*Vladimir Karas (Academy of Sciences of the Czech Rep.)*
*Ildar Khabibullin (MPA, Germany; IKI, Russia)*
*Henric Krawczynski (Washington Univ. in St. Louis, U.S.A.)*
*Julien Malzac (IRAP, Toulouse, France)*
*Frederic Marin (Obs. Strasbourg, France)*
*Herman Marshall (M.I.T., Cambridge, U.S.A.)*
*Giorgio Matt (Univ. Roma Tre, Italy)*
*Fabio Muleri (INAF, IAPS, Italy)*
*Carole Mundell (Univ. of Bath, U.K.)*
*Mark Pearce (KTH, Stockholm, Sweden)*
*Pierre-Olivier Petrucci (IPAG, Grenoble, France)*
*Juri Poutanen (Univ. Turku, Finland)*
*Roger Romani (Stanford Univ., U.S.A.)*
*Andrea Santangelo (Univ. Tuebingen, Germany)*
*Paolo Soffitta (INAF, IAPS, Italy)*
*Gianpiero Tagliaferri (INAF, Oss. Merate, Italy)*
*Roberto Taverna (Univ. Roma Tre, Italy)*
*Roberto Turolla (Univ. Padova, Italy)*
*Jacco Vink (Univ. of Amsterdam, The Netherlands)*
*Silvia Zane (Univ. College London, U.K.)*